\documentclass[aps,prb,twocolumn,,showpacs,floatfix,groupedaddress,superscriptaddress]{revtex4}
\usepackage{graphicx}

\begin {document}

\title{Non-Markovian reduced dynamics and entanglement evolution of two coupled
spins in a quantum spin environment}
% Force line breaks with \\

\author{Xiao-Zhong Yuan}
\affiliation{Department of Physics and Center for Theoretical Sciences, National Taiwan University,\\
 Taipei 10617, Taiwan
}
\affiliation{Department of Physics, Shanghai Jiao Tong University,
Shanghai 200240, China}
\author{Hsi-Sheng Goan}
\email{goan@phys.ntu.edu.tw}
\affiliation{Department of Physics and Center for Theoretical Sciences, National Taiwan University,\\
 Taipei 10617, Taiwan
}%
\author{Ka-Di Zhu}%
\affiliation{Department of Physics, Shanghai Jiao Tong University,
Shanghai 200240, China}
%\affiliation{

\date{September 12, 2006}

\begin{abstract}
The exact quantum dynamics of the reduced density matrix of
two coupled spin qubits in a quantum Heisenberg $XY$
spin star environment in the thermodynamic limit at arbitrarily finite
temperatures is obtained using a novel operator technique.
In this approach, the transformed Hamiltonian
becomes effectively Jaynes-Cumming
like and thus the analysis is also relevant to cavity quantum electrodynamics.
This special operator technique is mathematically simple and
physically clear, and allows us to treat
systems and
environments that could all be strongly coupled mutually and internally.
To study their entanglement evolution,
the concurrence of the reduced density matrix of the
two coupled central spins is also obtained exactly.
It is shown that the dynamics of the entanglement
depends on the initial state of the system and the coupling
strength between the two coupled central spins, the thermal temperature of the
spin environment and the interaction between the constituents of the
spin environment. We also investigate the effect of detuning which in
our model can be controlled by the strength of a locally applied
external magnetic field. It is found that the detuning
has a significant effect on the entanglement generation between the
two spin qubits.
\end{abstract}

%\keywords{Quantum spin models, entanglement, Concurrence, cavity QED}

\maketitle

\section{Introduction}
One of the most promising candidates for quantum computation is
spin systems \cite{1,kane,2,vrijen,3,skinner, ladd,sousa, friesen}
due to their long decoherence and relaxation time . Combined with
nanostructure technology, they have the potential advantage to
scale up to large systems. Just as other quantum systems, the spin
systems are inevitably influenced by their environment, especially
the spin environment. As a result, decoherence due to the presence
of the environment will cause the transition of a system from pure
quantum states to mixed ones. The decoherent behavior of a single
spin or several spins interacting with a spin bath has attracted
much attention in recent years \cite{4,5,6,Schliemann03}. The
interaction in such a spin bath system often leads to strong
non-Markovian behavior. The usual Markovian quantum master
equations which are widely used in the area of atomic physics and
quantum optics may fail for many spin bath models. Therefore it
becomes more and more important to develop methods that are
capable of going beyond the Markovian approximation
\cite{7,Breuerb,Hamdouni06}.

Entanglement has been recognized as one of the most amazing
aspects of quantum mechanics. It has been considered as important
resources for applications in quantum communication and
information processing, such as quantum teleportation
\cite{Bennett1}, quantum cryptography \cite{16}, quantum dense
coding \cite{17}, and telecloning \cite{18}. It is also believed
to be one of the features that make quantum computers more
powerful than classical ones. For spin systems, much attention has
been dedicated to the problem of thermal entanglement
\cite{21,Osborne02,20}, i.e., to quantify entanglement arising in
spin chains at thermal equilibrium with an environment or a
reservoir. In this approach, a thermal distribution of the system
energy levels is determined by the environment temperature, but
the detailed interaction between the system and environment, and
the evolution of the system toward the thermal equilibrium are
explicitly ignored.

A quantum system exposed to environmental modes is described by
the reduced density matrix when the environment modes are traced
over. The time evolution of the reduced density matrix is usually
very difficulty to obtain in the case of non-Markovian process.
Recently, the dynamics of the reduced density matrix for one-,
two- and three-spin-qubit systems in a spin bath described by the
transverse Ising model has been analyzed without making the
Markovian approximation, but using a perturbative expansion method
\cite{Paganelli02} or a mean-field approximation \cite{22,Ma05}.
The interaction between the system and the spin bath for these
cases was assumed to be of a Ising type. It has also been reported
recently that the exact reduced dynamics and for one- and
two-spin-qubit systems in a spin-star environment \cite{23} has
been derived and analyzed \cite{7,Breuerb}. There, the interaction
between the system and environment was assumed to be of a
Heisenberg $XY$ interaction.

In this paper, we study a two-spin-qubit system in a spin star
configuration, similar to the case studied in Ref.~\onlinecite{Hamdouni06}.
There are however several important differences between our model and
that of Ref.~\onlinecite{Hamdouni06}.
In Ref.~\onlinecite{Hamdouni06}, the interaction between the two-spin qubits
and the interaction between the constituents of
the spin environment are neglected, i.e., no internal dynamics
for both the spin
qubit system and the spin environment is considered.
In addition, the spin environment is assumed to be initially in an unpolarized
infinite temperature state. As a result, no dependence of the
environment temperature on the dynamics and entanglement is present.
It is under these conditions that the ``exact'' dynamics is reported in
Ref.~\onlinecite{Hamdouni06}.
Neglecting the
direct interaction among the constituents in the environment and
considering only the infinite temperature initial state may not
be proper in dealing with spin baths.
In this paper, we investigate a more general case.
Using a novel operator technique, we present an exact calculation of the
dynamics of the reduced density matrix of two coupled spins
interacting with a thermal spin bath at finite temperatures in the
thermodynamic limit.
In our model, the interaction
between the constituents of the spin environment, the
interaction between the two spin qubits, and
the interaction between the the spin qubit system and the spin environment
are all of the Heisenberg $XY$ type and can all be taken into account
simultaneously. In addition, we include also
the Zeeman coupling between the spin qubits and a locally applied
external magnetic field.
%The internal dynamics of the bath are
%explicitly taken into consideration.
To quantify quantum entanglement dynamics of the two coupled spins
under the influence of
the spin bath at arbitrarily finite temperatures in the thermodynamics
limit, we calculate the exact time
evolution of their concurrence.
Our model involves the Heisenberg $XY$ coupling
which has extensive applications for various quantum information
processing proposals \cite{9,10,11,12,13,14,Blais04,Sarovar05}.
In addition, the transformed Hamiltonian of the total system in our approach
becomes effectively Jaynes-Cumming like
and thus our analysis is also very relevant to
cavity quantum electrodynamics \cite{9,10,Blais04,Sarovar05}.

The paper is organized as follows. In Sec. II the model
Hamiltonian is introduced and the operator technique is employed
to obtain the reduced density matrix, taking into account the
memory effect of the environment. From the reduced density matrix,
the entanglement measure of concurrence of the coupled spin system is
calculated in Sec. III. Conclusions are given in Sec. IV.

\section{Model and Calculations}

 We consider a two-spin-qubit system interacting with  bath
spins via a Heisenberg \textit{XY} interaction. The system and
bath are composed of spin-$\frac{1}{2}$ atoms. We restrict
ourselves to a star-like configuration with coupling of equal
strength, similar to the cases considered in
Refs.~\onlinecite{7,Breuerb,23}. The interactions between bath
spins are also of \textit{XY} type. In
Refs.~\onlinecite{Paganelli02,22,Ma05}, a similar but somewhat
different type of Ising interactions between the constituents of
the spin bath was considered.
%The quantum Heisenberg
%$XY$ model has great applications in the regime of quantum
%computation and information processing \cite{9,10,11,12,13,14}.
The Hamiltonian for the total system is
\begin{eqnarray}
H&=&H_S+H_{SB}+H_B.
\end{eqnarray}
Here, $H_S$ and $H_B$ are the Hamiltonians of the system and bath
respectively, and $H_{SB}$ is the interaction between them
\cite{Breuerb,23}. They can be written as
\begin{eqnarray}
H_S&=&\mu_0(S_{01}^z+S_{02}^z)+\Omega\left(S_{01}^+S_{02}^-+S_{01}^-S_{02}^+\right),\\
H_{SB}&=&\frac{g_0}{\sqrt{N}}\left[\left(S_{01}^++S_{02}^+\right)
\sum_{i=1}^N S_i^-
 +\left(S_{01}^-+S_{02}^-\right) \sum_{i=1}^NS_i^+\right],\\
H_B&=&\frac{g}{N}\sum_{i\neq j}^N\left(S_i^+S_j^-+S_i^-S_j^+
\right),
\end{eqnarray}
where $\mu_0$ represents the coupling constant between a locally
applied external magnetic field in the $z$ direction and the spin
qubit system. $\Omega$ is the the coupling constant between two
qubit spins. $S_{0i}^+$ and $S_{0i}^-$ ($i$=1,2) are the spin-flip
operators of the qubit system spins, respectively. $S_i^+$ and
$S_i^-$ are the corresponding operators of the $i$th atom spin in
the bath. The indices of the sums for the spin bath run from $1$
to $N$, where $N$ is the number of the bath atoms. $g_0$ is the
coupling constant between the qubit system spins and bath spins,
whereas $g$ is that between the bath spins. Both coupling
strengths are rescaled such that the free energy is extensive and
a nontrivial finite limit of  $N\rightarrow\infty$ exists
\cite{Breuerb,24}.

By using collective angular momentum operators
$J_{\pm}=\sum_{i=1}^NS_{i}^\pm$, we rewrite the Hamiltonians,
Eqs.~(3) and (4), as
\begin{eqnarray}
H_{SB}&=&\frac{g_0}{\sqrt{2j}}\left[\left(S_{01}^++S_{02}^+\right) J_-+ \left(S_{01}^-+S_{02}^-\right)J_+\right],\\
H_B&=&\frac{g}{2j}\left(J_+J_-+J_-J_+ \right)-g,
\end{eqnarray}
where $j=N/2$ is the length of the pseudo-spin. After the
Holstein-Primakoff transformation \cite{25},
\begin{eqnarray}
J_+&=&b^+\left(\sqrt{2j-b^+b}\right), \hspace*{2mm}
J_-=\left(\sqrt{2j-b^+b}\right)b,
\end{eqnarray}
with $[b,b^+]=1$, the Hamiltonian, Eqs.~(5) and (6), can be
written as
\begin{eqnarray}
H_{SB}&=&g_0\left[\left(S_{01}^++S_{02}^+\right)\sqrt{1-\frac{b^+b}{N}}b+\left(S_{01}^-+S_{02}^-\right) b^+\sqrt{1-\frac{b^+b}{N}}\right],\\
H_B&=&g\left[ b^+\left(1-\frac{b^+b}{N}\right)
b+\sqrt{1-\frac{b^+b}{N}}bb^+\sqrt{1-\frac{b^+b}{N}}\right]-g.
\end{eqnarray}
In the thermodynamic limit (i.e. $N\longrightarrow\infty$) at
finite temperatures, we then have
\begin{eqnarray}
H_{SB}&=&g_0\left[\left(S_{01}^++S_{02}^+\right) b+\left(S_{01}^-+S_{02}^-\right) b^+\right],\\
H_B&=&2g b^+b.
\end{eqnarray}
Equations (2), (10) and (11) are then effectively equivalent to
the Hamiltonian of a Jaynes-Cumming type. They describe two
coupled qubits interacting with a single-mode thermal bosonic bath
field, so the analysis of the problem is also relevant to cavity
quantum electrodynamics quantum information processing proposals
\cite{9,10,Blais04,Sarovar05}. We note here that due to the high
symmetry of our model, the coupling to the environment is actually 
represented by a coupling to a single collective environment spin. 
After the Holstein-Primakoff transformation and in the thermodynamic limit,
this collective environment spin is transformed into a single-mode
bosonic thermal field. The effect of this single-mode environment on the
dynamics of the two coupled qubits is extremely non-Markovian.
This reflects onto, for example, the revival behavior of the
reduced density matrix or entanglement evolution of the two coupled
spins, which will be shown later.  
This is different from the usual environment models which consist of
very large degrees of freedom (e.g. many bosonic modes) and often
cause the reduced dynamics of the system of interest displaying an
exponential decay 
in time behavior. So the Markovian approximation usually used in
quantum optics master equation will not work in our model. One may
perform perturbation theory for weak-coupling case, but the
single-mode environment in our model will not remain in thermal
equilibrium state as usually assumed for an environment with
very large degrees of freedom in the weak-coupling master equation
approach.

Using a special operator technique, we can obtain the exact
reduced density matrix for the two coupled qubits by tracing over
the degrees of freedom of the bosonic bath at arbitrarily finite
temperatures. Reference \onlinecite{19} reported the theoretical
results of entanglement dynamics of a coupled two-level atoms
interacting with a cavity mode embedded in an effective atomic
environment. However the influence of the environmental
temperature was not considered. In Ref.~\onlinecite{26}, a
decoupled two-qubit system interacting with a single-mode thermal
field at resonance (i.e., zero detuning) in the context of cavity
electrodynamics was studied. There, the dynamics of the reduced
density matrix for the two-qubit system is obtained using the
method of the Kraus operator representation. In this paper, we use
a different approach of operator technique to obtain the exact
non-Markovian dynamics of the reduced density matrix for the
two-qubit system for our model of Eqs. (2)-(4) with the bath spins
in the thermodynamics limit, or equivalently Eqs. (2), (10) and
(11). Different from the case considered in Ref. \onlinecite{26},
our model furthermore includes the coupling between the two qubits
and investigates the effect of detuning (i.e., the single bosonic
bath mode is not necessarily resonant with the qubit transition
frequency). The detuning in our model is represented by
$(\mu_0-2g)$ and it could be controlled by the strength of a
locally applied magnetic field, i.e., the $\mu_0$ term in Eq.~(2).
We find that the detuning has a significant effect on the
entanglement generation between the two qubits.

We assume the initial density matrix of the total system to be
separable, i.e., $\rho(0)=|\psi\rangle\langle\psi|\otimes\rho_B$.
The density matrix
of the spin bath satisfies the Boltzmann distribution, that is
$\rho_B=e^{-H_B/T}/Z$, where $Z={\rm Tr}\left(e^{-H_B/T}\right)$
is the partition function, and
the Boltzmann constant has been set to one. At absolute zero
temperature, no excitation will exist. The bath is in a thoroughly
polarized state with all spins down. With the increase of
temperatures, the number of spin up atoms increases.
Note that in Ref.~\onlinecite{Hamdouni06}, the non-interacting bath
spins are assumed to be initially in the unpolarized infinite
temperature state.
The most general form of an initial pure state of the two-qubit
system is
\begin{eqnarray}
|\psi(0)\rangle=\alpha|00\rangle+\beta|11\rangle+\gamma|01\rangle+\delta|10\rangle,\\
\mbox{with}\quad |\alpha|^2+|\beta|^2+|\gamma|^2+|\delta|^2=1.
\end{eqnarray}
We might proceed the calculation with this general initial state, but
the final analytical solution would, however, be somewhat complicated.
For analytical simplicity, in the following we set
$\gamma=\delta=0$. We note that
%this restriction is not necessary.
the general initial qubit state case can be calculated in
a similar way presented below.

By taking the
initial state of the two qubit system to be
$|\psi\rangle=\alpha|00\rangle+\beta|11\rangle$, the reduced
density matrix can be written as
\begin{eqnarray}
\rho_s(t)
&=&\frac{1}{Z}|\alpha|^2\textrm{tr}_B\left[e^{-iHt}|00\rangle
e^{-H_B/T}\langle00| e^{iHt}\right]\nonumber\\
&&+\frac{1}{Z}\alpha\beta^*\textrm{tr}_B\left[e^{-iHt}|00\rangle
e^{-H_B/T}\langle11| e^{iHt}\right]\nonumber\\
&&+\frac{1}{Z}\alpha^*\beta\textrm{tr}_B\left[e^{-iHt}|11\rangle
e^{-H_B/T}\langle00| e^{iHt}\right]\nonumber\\
&&+\frac{1}{Z}|\beta|^2\textrm{tr}_B\left[e^{-iHt}|11\rangle
e^{-H_B/T}\langle11| e^{iHt}\right],
\end{eqnarray}
where
\begin{eqnarray}
Z=\frac{1}{1-e^{-2g/T}}.
\end{eqnarray}
The matrix $\rho_s(t)$ is a $4\times4$ matrix in the standard
basis $|00\rangle$, $|01\rangle$, $|10\rangle$, $|11\rangle$. In
order to obtain the exact reduced density matrix elements, we have
to evaluate Eq.~(14) exactly. However, it is difficult to do so
using usual methods, because the treated system and environment
could all be strongly coupled mutually and internally. In the
following, we will present a special operator technique to obtain
the exact density matrix elements. As shown below, our treatment
is mathematically simple and physically clear, and may be easily
extended to more complicated systems with strong coupling. Note
also that our method also applies to the case that the two-qubit
system is initially in a mixed state. For example, if the initial
state for the qubits is $\rho_s(0)=|\alpha|^2|00\rangle
\langle00|+|\beta|^2|11\rangle\langle11|$, the corresponding
reduced density matrix is Eq.~(14) provided that the second and
third terms on its right hand side are removed.

The basic idea of our operator technique is as follows. Before
tracing over the environmental degrees of freedom, we will first
convert the time evolution equation of the qubit system under the
action of the total Hamiltonian into a set of coupled
non-commuting operator variable equations. Then by introducing a
new set of transformation on the operator variables, we turn the
coupled non-commuting operator variable equations into commuting
ones. As a result, they can be solved exactly by using the general
method of solving coupled first-order ordinary differential
equations for ordinary variables. After that, the trace over the
environmental degrees of freedom can be performed and the exact
reduced dynamics of the qubit system can be obtained.

From the total Hamiltonian $H$, we can see that it consists of
operators $b$, $b^+$, $S_{0i}^-$ and $S_{0i}^+$, where $S_{0i}^-$,
and $S_{0i}^+$ change the $i$th ($i=1,2$) qubit spin from state
$|1\rangle_i$ to $|0\rangle_i$, and \textit{vice versa}.
%Since
%\begin{eqnarray}
%e^{-iHt}=1-iHt+\frac{(iHt)^2}{2!}\cdots,
%\end{eqnarray}
It is then obvious that we can write in a most general form that
\begin{eqnarray}
e^{-iHt}|11\rangle=A|00\rangle+B|01\rangle+C|10\rangle+D|11\rangle,
\end{eqnarray}
where $A$, $B$, $C$, and $D$ are functions of operators $b$,
$b^+$, and time $t$. Using the Schr\"{o}dinger equation identity
\begin{eqnarray}
i\frac{d}{dt}\left(e^{-iHt}|11\rangle\right)=H\left(e^{-iHt}|11\rangle\right)
\end{eqnarray}
and Eq. (16), we obtain
\begin{eqnarray}
\frac{d}{dt}A&=&-i\left[-\mu_0A+2gb^+bA+g_0b^+B+g_0b^+C\right],\\
\frac{d}{dt}B&=&-i\left[g_0bA+2gb^+bB+\Omega C+g_0b^+D\right],\\
\frac{d}{dt}C&=&-i\left[g_0bA+\Omega B+2gb^+bC+g_0b^+D\right],\\
\frac{d}{dt}D&=&-i\left[g_0bB+g_0bC+\mu_0D+2gb^+bD\right],
\end{eqnarray}
with initial conditions from Eq.~(16) being $A(0)=0$, $B(0)=0$,
$C(0)=0$, and $D(0)=1$. As $A$, $B$, $C$, $D$ are functions of
$b^+$ and $b$, they are operators and do not commute with each
other. Equations (18)-(21) are thus coupled differential equations
of non-commuting operator variables, which can not be solved by
using conventional methods for ordinary number variables.

The crucial observation to solve the problem is that the
Hamiltonian, Eqs.~(2), (10) and (11), is of an effective
Jaynes-Cumming type and it can be block-diagonalized in the
dressed state subspace of $|i,j;n\rangle$, with $i+j+n={\rm
constant}$. Here $|i,j\rangle$ represent the qubit states and
$|n\rangle$ are the bosonic field number states. As a result, we
may rewrite Eqs.~(18)-(21) in such a  subspace. By introducing the
following transformation
\begin{eqnarray}
A&=&b^+b^+e^{-i2g(b^+b+1)t}A_1,\\
B&=&b^+e^{-i2g(b^+b+1)t}B_1,\\
C&=&b^+e^{-i2g(b^+b+1)t}C_1,\\
D&=&e^{-i2g(b^+b+1)t}D_1,
\end{eqnarray}
equations (18)-(21) then become
\begin{eqnarray}
\frac{d}{dt}A_1&=&i(\mu_0-2g)A_1-ig_0\left(B_1+C_1\right),\\
\frac{d}{dt}B_1&=&-i\left[g_0(2+\hat{n})A_1+\Omega C_1+g_0D_1\right],\\
\frac{d}{dt}C_1&=&-i\left[g_0(2+\hat{n})A_1+\Omega B_1+g_0D_1\right],\\
\frac{d}{dt}D_1&=&-ig_0(1+\hat{n})\left(B_1+C_1\right)-i(\mu_0-2g)D_1,
\end{eqnarray}
where $\hat{n}=b^+b$. Note that for initial qubit state in
$|11\rangle$ on the left hand side of Eq.~(16), the
transformation, Eqs.~(22)-(25), is chosen in such a way that the
bosonic field operator(s) in front of the exponential term in
Eqs.~(22)-(25) together with its corresponding qubit state on the
right hand side of Eq.~(16) make $i+j+n$ a constant value and the
initial condition $D(0)=D_1(0)=1$. That is, the operator
coefficient $A$ of $|00\rangle$ state on the right hand side of
Eq.~(16) requires two field creation operators in Eq.~(22), $B$
and $C$ of $|01\rangle$ and $|10\rangle$ respectively require only
one field creation operator in Eqs.~(23) and (24), and $D$ of
$|11\rangle$ does not need any field operator in Eq.~(25). The
exponential term in Eqs.~(22)-(25) is introduced to make the
resultant equations more concise. As a consequence, the
coefficients of Eqs.~(26)-(29) after the transformation (22)-(25)
involve only the operator $\hat{n}$. Therefore $A_1$, $B_1$,
$C_1$, and $D_1$ are functions of $\hat{n}$ and $t$, and commute
with each other. We can then treat Eqs.~(26)-(29) as coupled
complex-number differential equations and solve them in a usual
way. This novel operators approach thus allows us to solve
Eq.~(16) and then consequently the non-Markovian dynamics of the
reduced density matrix of the qubit system.

We note again that the crucial point of the method used here is to find proper
transformations to change the coupled differential equations of
non-commuting operator variables to the coupled differential
equations of complex-number
variables. This can be done when the effective Hamiltonian can be
block-diagonalized. This is the case of Jaynes-Cumming model and other
models which contain interaction Hamiltonian of the forms of, for
example,
$\left(S_{01}^++S_{02}^+\right) bb+\left(S_{01}^-+S_{02}^-\right)
b^+b^+$, $S_{01}^+S_{02}^+b+S_{01}^-S_{02}^-b^+$, etc., regardless
how strong their interaction strengths are. If the total
effective Hamiltonian can not be block-diagonalized, for example,
for the effective spin-boson model in Ref.~\onlinecite{Irish}, the operator
method used here will then not apply to solve the problem exactly.

As we are working in the Schr\"{o}dinger picture, the basic
operator $\hat{n}=b^+b$ is time independent (sometimes the
operators could have time dependence explicitly; however this is
not the case here). From Eq.~(16) and Eqs.~(22)-(25), the initial
conditions at $t=0$ are given by
\begin{eqnarray}
A_1(0)&=&0,\\
B_1(0)&=&0,\\
C_1(0)&=&0,\\
D_1(0)&=&1.
\end{eqnarray}
In general, we can solve Eqs.~(26)-(29) exactly via the initial
conditions Eqs. (30)-(33). As we aim to obtain analytical
expressions for the reduced qubit dynamics, for the sake of
analytical simplicity we consider the on-resonant case, i.e.,
$\mu_0=2g$. We can easily tune the locally applied external
magnetic field to satisfy this condition. We will give the
numerical results for the off-resonant case in Fig.~6. We then
obtain for the on-resonant case
\begin{eqnarray}
A_1&=&\frac{-1}{3+2\hat{n}}+\frac{2g_0^2}{\sqrt{\Omega^2+8(3+2\hat{n})g_0^2}}\left[\frac{e^{i\lambda_1t}}{\lambda_1}-\frac{e^{i\lambda_2t}}{\lambda_2}\right],\\
B_1&=&C_1=-\frac{g_0}{\sqrt{\Omega^2+8(3+2\hat{n})g_0^2}}\left[e^{i\lambda_1t}-e^{i\lambda_2t}\right],\\
D_1&=&\frac{2+\hat{n}}{3+2\hat{n}}+\frac{2g_0^2(1+\hat{n})}{\sqrt{\Omega^2+8(3+2\hat{n})g_0^2}}\left[\frac{e^{i\lambda_1t}}{\lambda_1}-\frac{e^{i\lambda_2t}}{\lambda_2}\right],
\end{eqnarray}
where
\begin{eqnarray}
\lambda_{1,2}=\frac{-\Omega\pm\sqrt{\Omega^2+8(3+2\hat{n})g_0^2}}{2}.
\end{eqnarray}
Following the similar calculations above, we can evaluate the time
evolution for the initial two-qubit spin state of $|00\rangle$. Let
\begin{eqnarray}
e^{-iHt}|00\rangle=E|00\rangle+F|01\rangle+G|10\rangle+K|11\rangle.
\end{eqnarray}
In a similar way, we have
\begin{eqnarray}
E&=&e^{-i2g\left(b^+b-1\right)t}E_1,\\
F&=&be^{-i2g\left(b^+b-1\right)t}F_1,\\
G&=&be^{-i2g\left(b^+b-1\right)t}G_1,\\
K&=&bbe^{-i2g\left(b^+b-1\right)t}K_1,
\end{eqnarray}
and then obtain
\begin{eqnarray}
E_1&=&\frac{\hat{n}-1}{2\hat{n}-1}+\frac{2g_0^2\hat{n}}{\sqrt{\Omega^2+8(2\hat{n}-1)g_0^2}}\left[\frac{e^{i\lambda_1't}}{\lambda_1'}-\frac{e^{i\lambda_2't}}{\lambda_2'}\right],\\
F_1&=&G_1=-\frac{g_0}{\sqrt{\Omega^2+8(2\hat{n}-1)g_0^2}}\left[e^{i\lambda_1't}-e^{i\lambda_2't}\right],\\
K_1&=&\frac{-1}{2\hat{n}-1}+\frac{2g_0^2}{\sqrt{\Omega^2+8(2\hat{n}-1)g_0^2}}\left[\frac{e^{i\lambda_1't}}{\lambda_1'}-\frac{e^{i\lambda_2't}}{\lambda_2'}\right],
\end{eqnarray}
where
\begin{eqnarray}
\lambda_{1,2}'=\frac{-\Omega\pm\sqrt{\Omega^2+8(2\hat{n}-1)g_0^2}}{2}.
\end{eqnarray}
From Eq.~(14) and all the results that we obtained, the reduced
density matrix can be written as
\begin{eqnarray}
\rho_s(t)&=&\left(\begin{array}{cccc}
\rho_{11} & 0 &0&\rho_{14} \\
0&\rho_{22} & \rho_{23}&0\\
0&\rho_{32} & \rho_{33}&0\\
\rho_{14}^*&0&0&\rho_{44}\\
\end{array}\right),
\end{eqnarray}
where
\begin{eqnarray}
\rho_{11}&=&|\alpha|^2\frac{1}{Z}\sum_{n=0}^\infty
E_1E_1^+e^{-2gn/T}\nonumber\\
&&+|\beta|^2\frac{1}{Z}\sum_{n=0}^\infty
A_1A_1^+(n+1)(n+2)e^{-2gn/T},\\
\rho_{14}&=&\alpha\beta^*\frac{1}{Z}\sum_{n=0}^\infty
E_1D_1^+e^{-2gn/T}e^{i4gt},\\
\rho_{22}&=&\rho_{23}=\rho_{32}=\rho_{33}\nonumber\\
&=&|\alpha|^2\frac{1}{Z}\sum_{n=1}^\infty
F_1F_1^+ne^{-2gn/T}\nonumber\\
&&+|\beta|^2\frac{1}{Z}\sum_{n=0}^\infty B_1B_1^+(n+1)e^{-2gn/T},\\
\rho_{44}&=&|\alpha|^2\frac{1}{Z}\sum_{n=2}^\infty
K_1K_1^+n(n-1)e^{-2gn/T}\nonumber\\
&&+|\beta|^2\frac{1}{Z}\sum_{n=0}^\infty
D_1D_1^+e^{-2gn/T}.
\end{eqnarray}
In Eqs.~(48)-(51), the trace over the environmental degrees of
freedom has been performed and the operator $\hat{n}$ has been
replaced by its eigenvalue $n$. In a similar way, the solutions
for the reduced dynamics of the two-coupled spins under the
influence of the quantum Heisenberg $XY$ spin star bath in the
thermal dynamics limit at arbitrarily finite temperatures for
arbitrary initial states of
$|\psi(0)\rangle=\alpha|00\rangle+\beta|11\rangle+\gamma|01\rangle+\delta|10\rangle$
can be obtained.

\section{Concurrence and Entanglement Dynamics}
We use the concurrence \cite{27} to measure the entanglement between the two
coupled qubit spins. It is defined as \cite{27}
\begin{eqnarray}
C_{12}=\textrm{max}\{\lambda_1-\lambda_2-\lambda_3-\lambda_4,0\},
\end{eqnarray}
where the quantities
$\lambda_1\geq\lambda_2\geq\lambda_3\geq\lambda_4$ are the square
roots of the eigenvalues of the operator
\begin{eqnarray}
R_{12}=\rho_s\left(\sigma^y\otimes\sigma^y\right)\rho_s^*\left(\sigma^y\otimes\sigma^y\right).
\end{eqnarray}
We find $\lambda_i$ are values in decreasing order of
$\sqrt{\rho_{11}\rho_{44}}+|\rho_{14}|$,
$|\sqrt{\rho_{11}\rho_{44}}-|\rho_{14}||$, $2\rho_{22}$, and $0$.
For the system with an initial state of $|\psi\rangle=|00\rangle$,
i.e., the both spins in the ground state, we plot the time
evolution of the concurrence in Fig.~1. Although there is no
initial entanglement and no coupling between the two spins, it is
interesting to notice that the entanglement between the two spins
after some time is present as shown in Fig.~1(a). This confirms
that the environment which usually causes the decoherence of the
system can nevertheless entangle qubits that are initially
prepared in a separable state \cite{26,28,29,30,Napoli}. This is
mainly due to the fact that the two spins are coupled to the same,
common environment which then in turn generates some effective
interaction between the two spins even if they were originally
decoupled. The result, however, depends on the environmental
temperature. Further numerical calculations show that no
entanglement is generated, for example, for $T>8g$. As the
coupling between the two qubit spins is switched on even though
the value is small, the ``collapse'' and ``revival'' of the
entanglement as a function of time are demonstrated in Fig.~1(b).
This is in analogy to the collapse and revival of atomic
population inversion of a single two-level atom interacting with a
single mode field initially in a coherent state \cite{31}, a Fock
state \cite{19} or a squeezed state \cite{Fleischhauer93} in
quantum optics. Here from the Hamiltonian, Eqs.~(2), (10) and
(11), this novel phenomenon of entanglement arises from
two-coupled qubits interacting with a single mode field initially
however in a thermal state. The reasons for causing the collapse
and revival behaviors in these cases could be similar, that is the
Rabi (or time evolution) oscillations associated with different
excitations have different frequencies . Consequently, as the time
increases, these Rabi (or time evolution) oscillations become
uncorrelated leading to a collapse behavior. As time is further
increased, the correlation is restored and the revival occurs.
%Also it is
%different from the results of Ref.~\onlinecite{19} where the cavity
%field is in the photon Fock state.

\begin{figure}[tb]
\includegraphics [width=8.2cm] {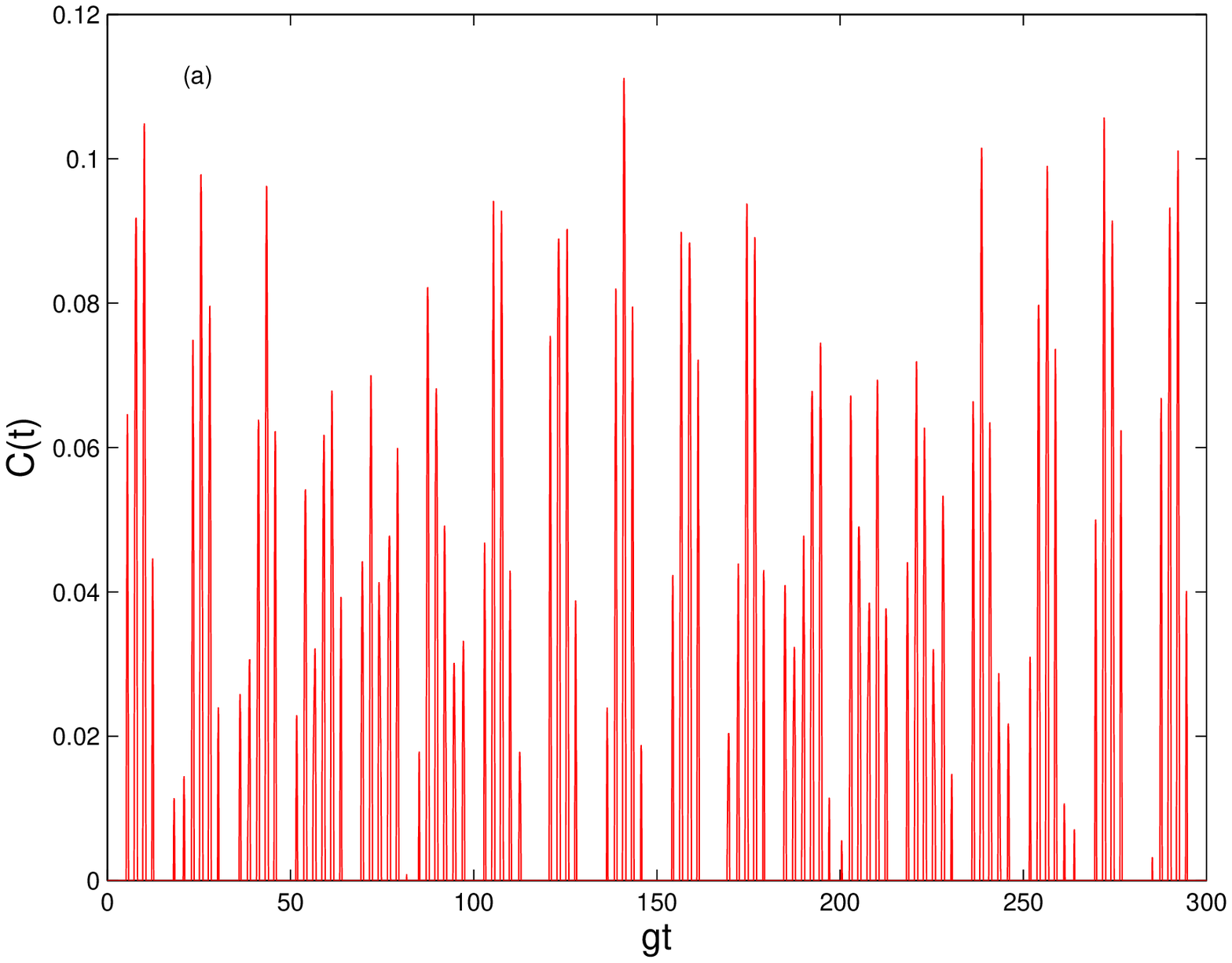}
\includegraphics [width=8.2cm] {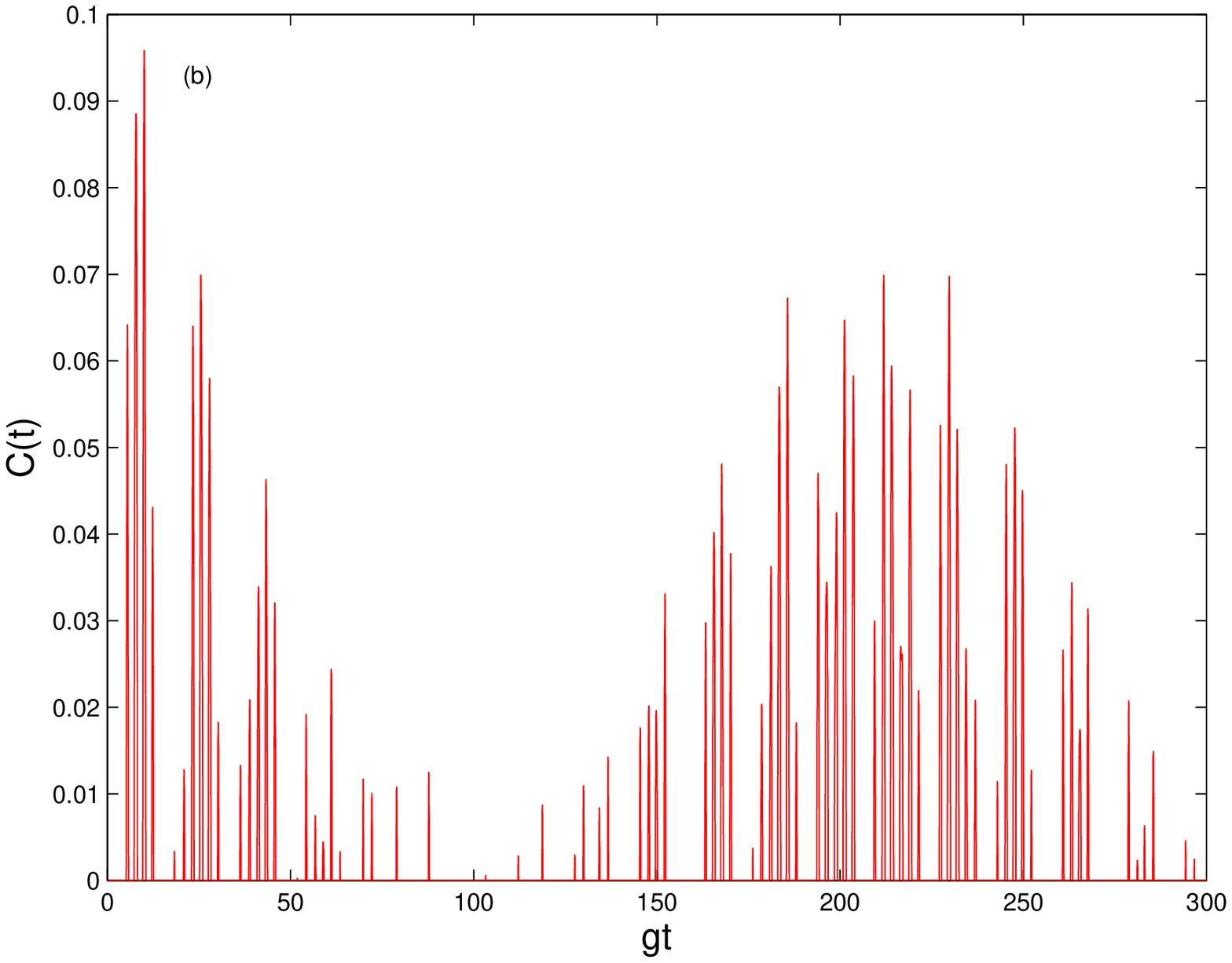}
\caption{(color online) Time evolution of concurrence for an initial two-qubit
state of $|\psi\rangle=|00\rangle$ for different values of
$\Omega$; (a) $\Omega=0$ and (b) $\Omega=0.03g$. Other parameters are
$\mu_0=2g$, $g_0=g$, $T=1g$.}
\end{figure}

Figure 2 shows the time evolution of the concurrence for the
system in the initial state of
$|\psi\rangle=\frac{1}{\sqrt{2}}(|00\rangle+|11\rangle)$, i.e., an
maximally entangled state. At high temperature, the state loses
its entanglement completely for a short period of time, and then
it is partially entangled again some times later. This is in
agreement with the results of Ref.~\onlinecite{Hamdouni06} where
an initially unpolarized infinitely temperature states of the spin
bath is assumed. However, our results are temperature dependent.
At a very low temperature, the concurrence exhibits the behavior
of oscillation between 1 and 0.35. With increasing temperatures,
the concurrence decreases more quickly and oscillates disorderly
in the lower value region. At a fixed time $t$, the concurrence
decreases with temperatures and a critical temperature $T_c$
exists, above which the entanglement vanishes. However, this
critical temperature $T_c$ is time-dependent and sensitive to the
initial state of the system. Figure 3 illustrates the time
evolutions of the concurrence for different values of the coupling
constant $g_0$. As expected, increasing the value of the coupling
constant has similar effects as increasing the value of the
environmental temperature, i.e., the decay rate of the concurrence
increases. Figure 4 presents the time evolution of the concurrence
for different inner-bath-spin coupling constants $g$. We see that
the concurrence increases with the increase of $g$. This confirms
that strong quantum correlations within the environment suppress
decoherence \cite{Paganelli02,Tessieri03,Dawson05} and thus
perhaps also disentanglement. As shown in the inset, the
concurrence is regained sometime later. However, it appears
disorderly without a particular pattern. Similar behaviors arise
for Fig.~2, Fig.~3, and Fig.~5 in the long time scales, which
reflect the Non-Markovian dynamics of the system. In Fig.~5, we
show the effect of the coupling between the two qubit spins on the
concurrence. It is obvious that the coupling benefits the
entanglement. Note that the initial state of the system in this
case is different from that in Fig.~1. So, if the system is
initially prepared in a maximally entangled state, the larger the
coupling constant $\Omega$ is, the more slowly the entanglement
decays.
\begin{figure}[tb]
\includegraphics  [width=8.2cm] {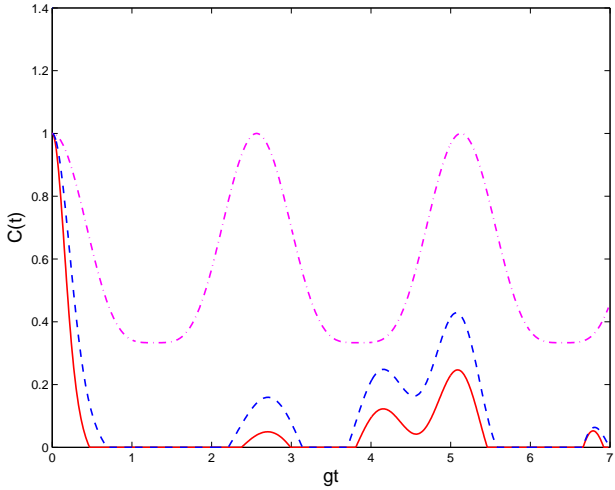}
\caption{(color online) Time evolution of concurrence for an initial two-qubit
state of $|\psi\rangle=\frac{1}{\sqrt{2}}(|00\rangle+|11\rangle)$
for different temperatures;
$T=10g$ (solid curve), $T=5g$ (dashed curve) and $T=0.1g$ (dot
dashed curve). Other parameters are  $\mu_0=2g$, $g_0=g$, $\Omega=0$.}
\end{figure}

\begin{figure}[tb]
\includegraphics  [width=8.2cm] {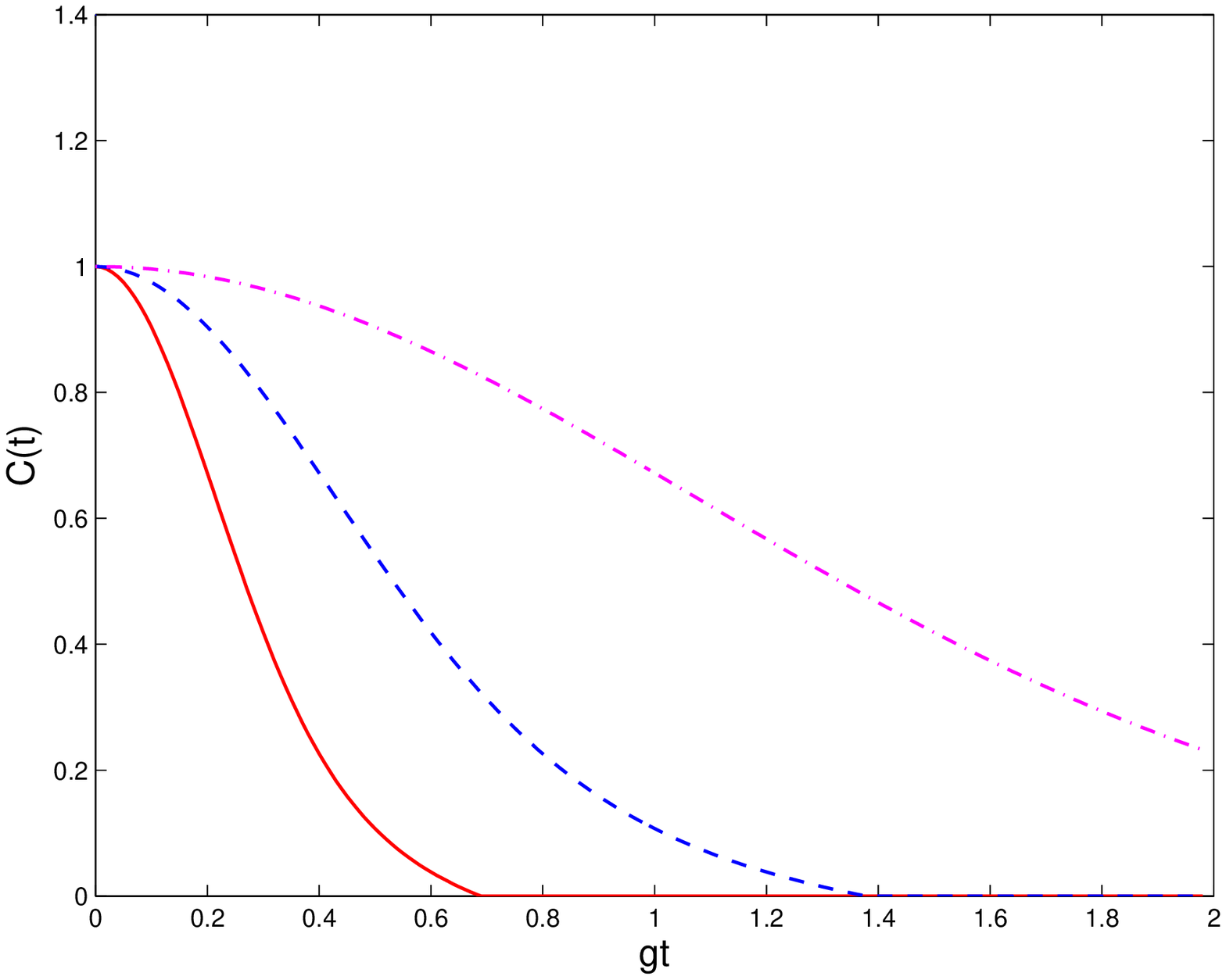}
\caption{(color online) Time evolution of concurrence for an initial two-qubit
state of $|\psi\rangle=\frac{1}{\sqrt{2}}(|00\rangle+|11\rangle)$
for different values of $g_0$;
$g_0=g$ (solid curve), $g_0=0.5g$ (dashed curve) and $g_0=0.2g$
(dot dashed curve).  Other parameters are $\mu_0=2g$, $T=5g$, $\Omega=0$.}
\end{figure}

\begin{figure}[tb]
\includegraphics  [width=8.2cm] {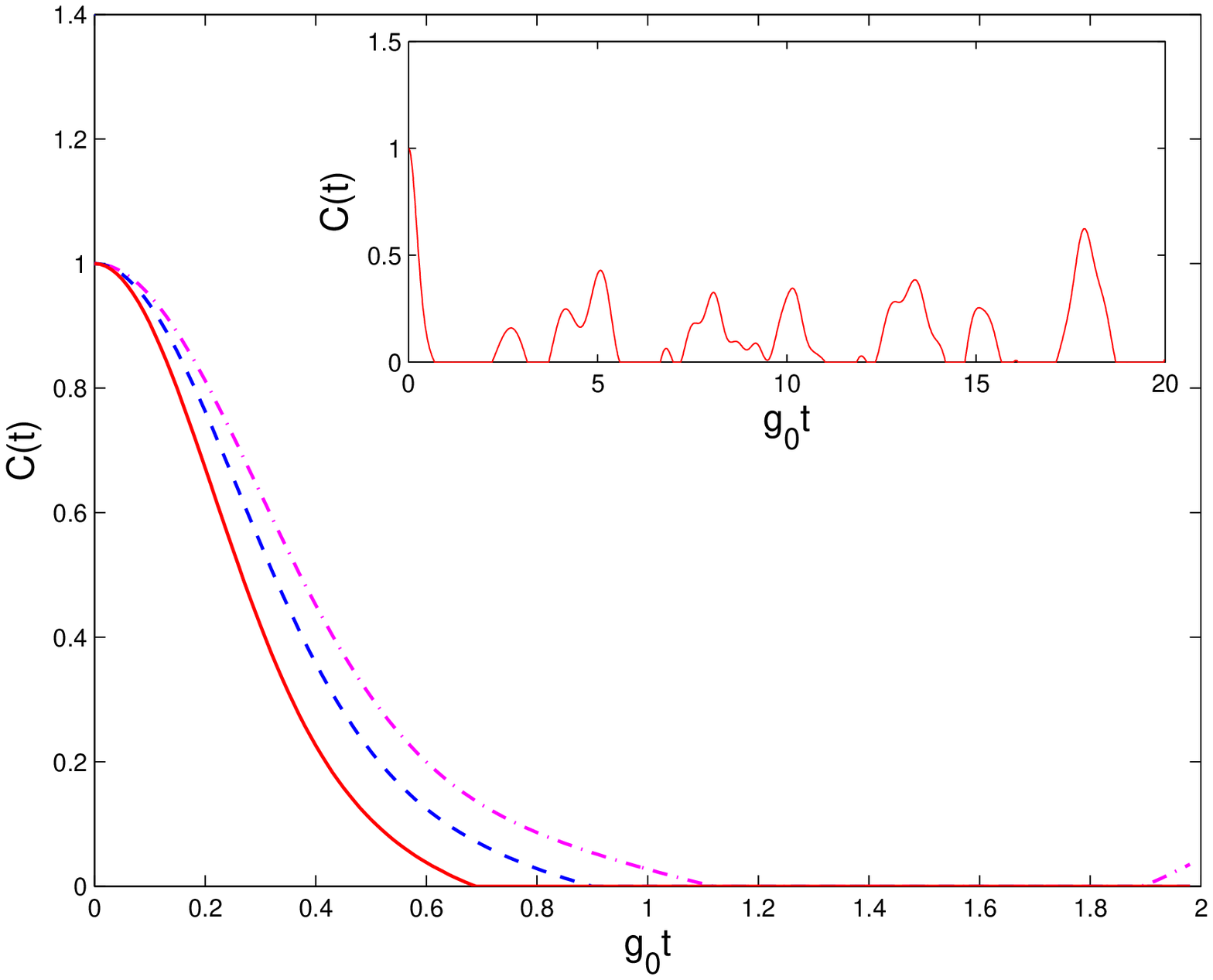}
\caption{(color online) Time evolution of concurrence for an initial two-qubit
state of $|\psi\rangle=\frac{1}{\sqrt{2}}(|00\rangle+|11\rangle)$
for different values of $g$;
$g=g_0$ (solid curve), $g=1.5g_0$ (dashed curve) and $g=2g_0$ (dot
dashed curve). Other parameters are  $\mu_0=2g$, $T=5g_0$, $\Omega=0$.
The inset shows the long time behavior of
concurrence for $g=g_0$.}
\end{figure}

\begin{figure}[tb]
\includegraphics  [width=8.2cm] {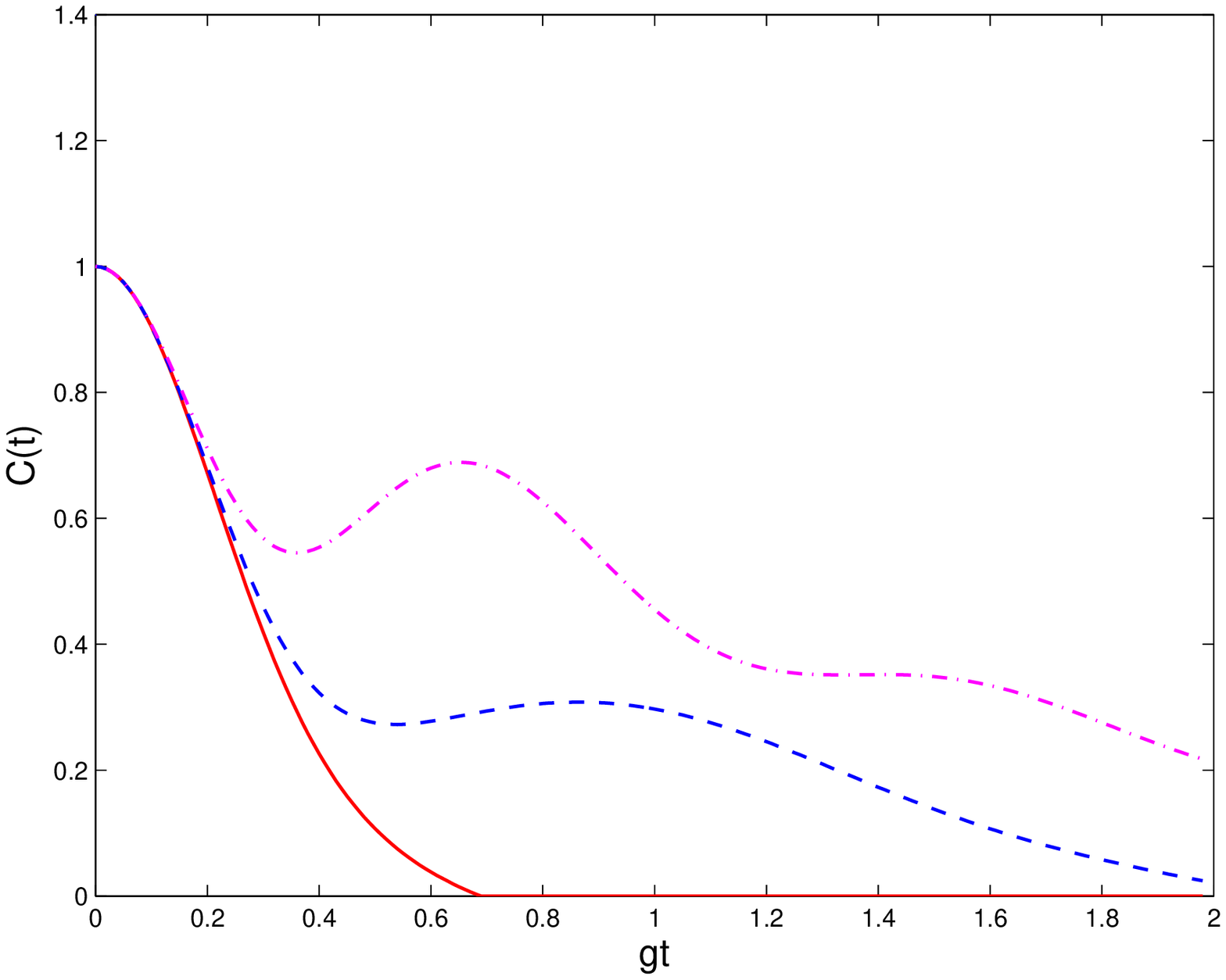}
\caption{(color online) Time evolution of concurrence for an initial two-qubit
state of $|\psi\rangle=\frac{1}{\sqrt{2}}(|00\rangle+|11\rangle)$
for different values of $\Omega$; $\Omega=0$ (solid curve),
$\Omega=3g$ (dashed curve), and
$\Omega=6g$ (dot dashed curve). Other parameters are $\mu_0=2g$,
$g_0=g$, $T=5g$.}
\end{figure}

\begin{figure}[tb]
\includegraphics  [width=8.2cm] {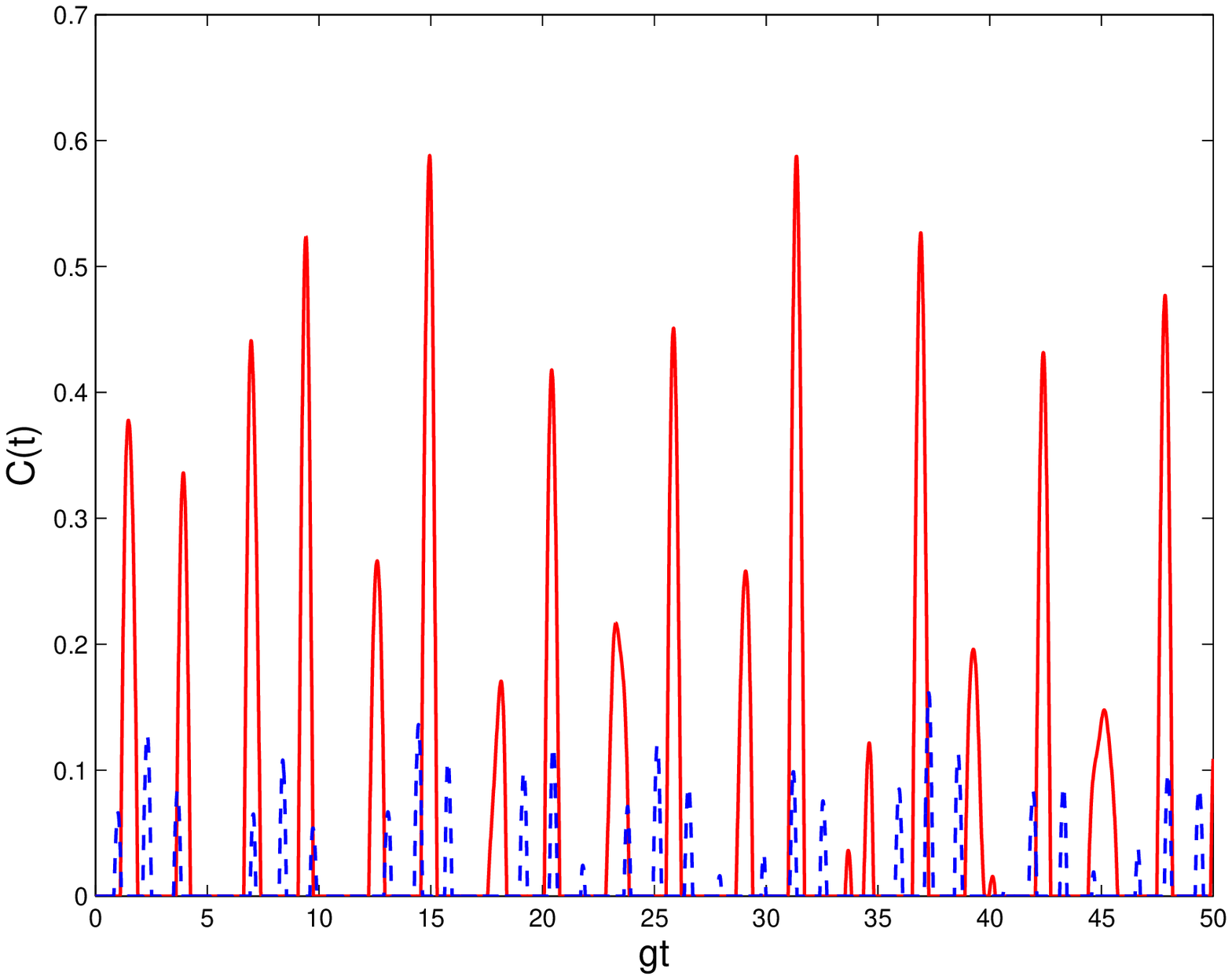}
\caption{(color online) Time evolution of concurrence for an initial two-qubit
state of $|\psi\rangle=|11\rangle$ for different values of detuning;
 $\mu_0=3.5g$ (solid curve), and
$\mu_0=6g$ (dashed curve).  Other parameters are $g_0=g$, $T=1g$, $\Omega=0$.}
\end{figure}

When the two spins are initially prepared in their excited state,
i.e. $|\psi\rangle=|11\rangle$, the result is quiet different from
that in Fig. 1. If the detuning $\mu_0=2g$, i.e., at the
on-resonant case, no entanglement between the two qubit spins
exists at any temperatures, even with a strong interaction
$\Omega$ between them. This is consistent with the result for two
qubit atoms obtained by Kim et al., \cite{26} where the coupling
between the two qubit atoms is not considered and the detuning
between the two atoms and the single-mode field is zero. If the
detuning $\mu_0\neq 2g$, i.e., in the off-resonant case, the two
qubit spins will entangle via the environment again as shown in
Fig.~6. So the entanglement generation of the two spins in this
case is very sensitive to the detuning which can be controlled in
our model by the locally applied external magnetic field.

\section{Conclusion}
 We have studied the exact entanglement evolution of two coupled
qubit spins in a model of a quantum Heisenberg $XY$ spin star
environment in the thermodynamic limit. The dynamics of the
reduced density matrix of the two coupled spins is analytically
obtained in terms of a novel operator technique which is
mathematically simple and physically clear. In our analysis, the
transformed Hamiltonian becomes effectively Jaynes-Cumming like
and thus the results are also relevant to cavity quantum
electrodynamics.

The time evolutions of the
concurrence of the two coupled spin qubits for different initial
conditions are evaluated exactly. The results show that the dynamics
of the entanglement strongly depends on the initial state of the
system, the coupling between the two spin qubits,
the interaction between the qubit system and the environment,
the interactions between the constituents of the spin environment,
the environmental temperatures, as well as the detuning controlled by a
locally applied external magnetic field. We have also found that
if the two coupled spin qubits are initially prepared in the
ground state, the entanglement between them will exhibit the
``collapse'' and ``revival'' behavior with time due to the
interaction between the two spin qubits and the environment, in
analogy to the collapse and revival of the atomic population inversion in
quantum optics.

\begin{acknowledgments}
X.Z.Y. and H.S.G. would like to acknowledge support from the National Science
Council, Taiwan, under grant numbers NSC94-2112-M-002-028,
NSC95-2112-M-002-018 and NSC95-2112-M-002-054. H.S.G. also thanks support
from the National Taiwan University under grant number 95R0034-02 and
is grateful to the National Center for High-performance Computing, Taiwan, for
computer time and facilities.
\end{acknowledgments}

%\newpage

%\centerline{\large{\bf The Figure Captions}}

\end{document}